\documentstyle[12pt]{article}
\begin{document}

\topmargin 0pt
\oddsidemargin=-0.4truecm
\evensidemargin=-0.4truecm

\renewcommand{\thefootnote}{\fnsymbol{footnote}}

\vspace*{-2.0cm}

\vspace*{0.5cm}
\begin{center}
  {\Large\bf Non-renormalization of  induced charges and 
constraints on strongly coupled theories.} \\
  \medskip S.~L.~Dubovsky,
   D.~S.~Gorbunov,
M.~V.~Libanov, and
V.~A.~Rubakov

  \medskip
  {\small
     Institute for Nuclear Research of
         the Russian Academy of Sciences,\\  60th October Anniversary
  Prospect, 7a, 117312 Moscow, 
Russia
  }
\end{center}

\begin{abstract}
It is shown that global fermionic charges induced in vacuum by slowly
varying, topologically non-trivial background scalar fields are not
renormalized provided that expansion in momenta of background fields
is valid.  This suggests that strongly coupled theories obey induced
charge matching conditions which are analogous, but generally not
equivalent, to 't Hooft anomaly matching conditions. We give a few
examples of induced charge matching. In particular, the corresponding
constraints in softly broken supersymmetric QCD suggest non-trivial
low energy mass pattern, in full accord with the results of direct
analyses.
\end{abstract}

\renewcommand{\thefootnote}{\arabic{footnote}}
\setcounter{footnote}{0}

\section{Introduction}
\label{1}

Four dimensional gauge theories strongly coupled at low energies often
exhibit interesting content of composite massless fermions. This
property is potentially important for constructing composite models of
quarks and leptons, which is long being considered as a possible
``major step on our way into the nature of matter''~\cite{Okun}.
Powerful constraints on the low energy spectrum are provided by t'
Hooft anomaly matching conditions~\cite{tHooft}. These are heavily
used, in particular, in establishing duality properties of
supersymmetric gauge theories (see, e.g., refs.~\cite{Intriligator,Poppitz}
and references therein).

The basis for anomaly matching is provided by the Adler--Bardeen
non-renormalization theorem~\cite{adler}. In non-Abelian theories, the
absence of radiative corrections to anomalies is intimately connected
to topology: if one introduces background gauge fields corresponding
to the flavor symmetry group, the anomalies in global currents are
proportional to the topological charge densities of these background
fields. Integer-valuedness of global fermionic charges, on the one
hand, and integer-valuedness of topological charges of background
gauge fields, on the other, imply that anomaly equations should not
get renormalized.

Gauge field backgrounds are not the only ones that may have
topological properties. Topology is inherent also in scalar background
fields of Skyrmion type. Indeed, one loop calculations~\cite{GW} show
that slowly varying in space, static scalar fields induce, in vacuum,
fermionic global charges which are proportional to the topological
charges of the background. By analogy to triangle anomalies, this
suggests that induced charges do not receive radiative corrections,
and hence may serve as constraints on low energy spectrum of strongly 
coupled theories \cite{Rubakov}. Unlike triangle anomalies, however, 
the one-loop
expressions for the induced charges are promoted to full quantum
theory only if the expansion in momenta of the background fields is
valid in the full theory. The latter property can often be established
to all orders of perturbation theory (exceptions are easy to
understand); in some models the validity of the expansion in momenta
can be also shown non-perturbatively.

We will see that induced charge matching conditions emerging in this
way have a certain relation to anomaly matching.  However, in some
cases the two sets of matching conditions are inequivalent, so the
induced charges give additional information on the properties of
the low energy theory. This information is particularly interesting
in softly broken supersymmetric gauge theories.

Fermionic charges in vacuum are induced due to Yukawa interactions of
fermions with background scalar fields.  These interactions introduce
masses to some of the fermions in the fundamental theory and hence
explicitly break a subgroup of the flavor group. As a consequence,
some of the fermions of the low energy effective theory acquire
masses. Induced charge matching conditions constrain the resulting
mass pattern of the effective theory. We will see that these
conditions are satisfied automatically (provided the triangle
anomalies match) if {\it all} composite fermions charged under
explicitly broken flavor subgroup become massive. The latter situation
is very appealing intuitively; however, we are not aware of  any argument
implying that it should be generic.

This paper is organized as follows. In section \ref{2} we show that
global charges induced in vacuum by slowly varying background scalar
fields do not get renormalized provided that the derivative expansion 
is valid. In section \ref{2a} we discuss exceptional situations by
presenting a model where the derivative expansion fails already at the
one loop level. In section \ref{3} we give several examples of induced
charge matching (ordinary QCD, supersymmetric $N=1$ QCD exhibiting the
Seiberg duality~\cite{Seiberg-du}, SQCD with softly broken
supersymmetry). We conclude in section \ref{4} by discussing the
relation between induced charges and triangle anomalies.

\section{Non-renormalization of induced charges}
\label{2}

To be specific, let us consider QCD with $N_c$ colors and $N_f$
massless fermion flavors.  Let $\psi^a$ and
$\tilde{\psi}_{\tilde{a}}$, $a,\tilde{a}=1,\ldots,N_f$, denote
left-handed quarks and anti-quarks, respectively. To probe this
theory, we introduce background scalar fields
$m^{\tilde{a}}_{b}(x)$ of the following form,
\[
m^{\tilde{a}}_{b}(x)=m_0U^{\tilde{a}}_{b}(x)\;,
\]
where $m_0$ is a constant and $U$ is an $SU(N_f)$ matrix at each point
$x$.  
Let these fields interact with quarks and anti-quarks,
\begin{equation}
{\cal L}_{\rm int}=\tilde{\psi}_{\tilde{a}}m^{\tilde{a}}_b\psi^b+{\rm h.c.}
\label{V*}
\end{equation}
Besides the global $SU(N_f)_L\times SU(N_f)_R$ symmetry, the theory
exhibits non-anomalous baryon symmetry, $\psi\to{\rm
e}^{i\alpha}\psi$, $\tilde{\psi}\to{\rm e}^{-i\alpha}\tilde{\psi}$,
$m\to m$. The baryonic current is conserved and obtains non-vanishing
vacuum expectation value in the presence of the background scalar fields.

\begin{figure}[t]
\begin{center}
\begin{picture}(0,0)%
\includegraphics{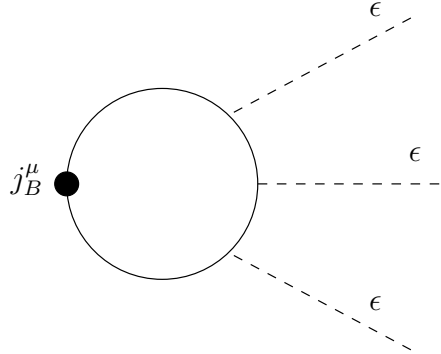}%
\end{picture}%
\setlength{\unitlength}{3947sp}%
\begingroup\makeatletter\ifx\SetFigFont\undefined
\def\x#1#2#3#4#5#6#7\relax{\def\x{#1#2#3#4#5#6}}%
\expandafter\x\fmtname xxxxxx\relax \def\y{splain}%
\ifx\x\y   
\gdef\SetFigFont#1#2#3{%
  \ifnum #1<17\tiny\else \ifnum #1<20\small\else
  \ifnum #1<24\normalsize\else \ifnum #1<29\large\else
  \ifnum #1<34\Large\else \ifnum #1<41\LARGE\else
     \huge\fi\fi\fi\fi\fi\fi
  \csname #3\endcsname}%
\else
\gdef\SetFigFont#1#2#3{\begingroup
  \count@#1\relax \ifnum 25<\count@\count@25\fi
  \def\x{\endgroup\@setsize\SetFigFont{#2pt}}%
  \expandafter\x
    \csname \romannumeral\the\count@ pt\expandafter\endcsname
    \csname @\romannumeral\the\count@ pt\endcsname
  \csname #3\endcsname}%
\fi
\fi\endgroup
\begin{picture}(2862,2124)(151,-1423)
\put(2751,-211){\makebox(0,0)[lb]{\smash{\SetFigFont{12}{14.4}{rm}$\epsilon$}}}
\put(251,-380){\makebox(0,0)[lb]{\smash{\SetFigFont{12}{14.4}{rm}$j_B^\mu$}}}   
\put(2501,700){\makebox(0,0)[lb]{\smash{\SetFigFont{12}{14.4}{rm}$\epsilon$}}}
\put(2501,-1150){\makebox(0,0)[lb]{\smash{\SetFigFont{12}{14.4}{rm}$\epsilon$}}}
\end{picture}
\caption{Leading order contribution to the induced baryonic current.
\label{threetails}}
\end{center}
\end{figure}

To the leading order in momenta, the one-loop expression for this 
induced current is~\cite{GW}
\begin{equation}
\label{complete}
\langle j_B^\mu\rangle = \frac{N_c}{24 \pi^2}  \epsilon^{\mu\nu\lambda\rho}
           \mbox{Tr} \left( U\partial_{\nu} U^{\dagger}
                 U\partial_{\lambda} U^{\dagger}  U\partial_{\rho}
            U^{\dagger}        
            \right)\;.
\label{*}
\end{equation}
This expression can be obtained by considering a configuration which
in the vicinity of a given point $x$ has the following form,
\begin{equation}
U(x)={\bf 1} + \epsilon (x)\;,
\label{VI*}
\end{equation} 
 where $\epsilon (x)$ is a small and slowly varying anti-Hermitean
background field.  To the leading order in momenta, one-loop induced
baryonic current is then given by the diagram of Fig.~\ref{threetails}
with fermions of mass 
$m_0$ running in the loop. The complete expression~(\ref{complete}) is 
reconstructed by making use of $SU(N_f)_L\times SU(N_f)_R$ global symmetry.     

A remarkable property of eq. (\ref{*}) is that the baryonic charge induced
in vacuum by slowly varying, time-independent background field $U({\bf x})$
with 
\[
U({\bf x})\to {\bf 1}\ \  {\rm as} \ \ |{\bf x}|\to \infty
\]
is proportional to the topological number of the background,
\begin{equation}
\langle B\rangle =N_cN[U]\;,
\label{**}
\end{equation}
where
\[
N[U]=\frac{1}{24 \pi^2} \int\!d^3 x~ \epsilon^{ijk}
           \mbox{Tr} \left( U\partial_i U^{\dagger}
                 U\partial_j U^{\dagger}  U\partial_k U^{\dagger}        
            \right)\;,
\]
The higher derivative terms  omitted in eq. (\ref{*}) do not
contribute to $\langle B\rangle$.

Let us see that eq. (\ref{**}) does not get renormalized in the full quantum
theory provided the expansion in momenta of the background field works.
Let us consider the same theory with the gauge coupling $\alpha$ depending
on coordinates $x$. The induced current is now a functional of $\alpha(y)$
and $U(y)$,
\[
\langle j_B^\mu(x)\rangle=j^{\mu}[x;\alpha(y);U(y)]\;.
\]
At slowly varying $\alpha(y)$ we expand $\langle j_B^\mu(x)\rangle$ in
derivatives of $\alpha$ at the point $x$,
\begin{equation}
\label{***}
\langle j_B^\mu(x)\rangle=
{\cal J}^\mu\left [x;\alpha(x);U(y)\right]
+B^{\mu\nu}[x;\alpha(x);U(y)]\partial_\nu\alpha(x)+
O\left[(\partial\alpha)^2\right]\;,
\end{equation}
where the coefficients on the right hand side are now functions,
rather than functionals, of $\alpha(x)$ (but still functionals of $U(y)$).
The structure analogous to the right hand side of eq.~(\ref{*}) appears
in the derivative expansion of the first term on the right hand side of
eq. (\ref{***}),
\[
{\cal J}^\mu= \frac{f(\alpha)}{24 \pi^2} \epsilon^{\mu \nu\lambda\rho}
\mbox{Tr} \left( U\partial_\nu U^{\dagger} U\partial_\lambda
U^{\dagger} U\partial_\rho U^{\dagger} \right)+\ldots
\]
Our purpose is to show that $f(\alpha)$ is independent of $\alpha$, and
hence $f(\alpha)=N_c$.

Let us make use of the conservation of $\langle j_B^\mu(x)\rangle$,
\begin{equation}
\partial_\mu\langle j_B^\mu(x)\rangle=0\;.
\label{+}
\end{equation}
If $f$ were a non-trivial function of $\alpha$, the divergence of
${\cal J}^\mu$ would contain the term
\[
\frac{1}{24 \pi^2}  \epsilon^{\mu \nu\lambda\rho}
           \mbox{Tr} \left( U\partial_\nu U^{\dagger}
                 U\partial_\lambda U^{\dagger}  U\partial_\rho U^{\dagger}        
            \right)\frac{\partial f}{\partial \alpha}\partial_\mu\alpha\;.
\]
The only possible source of cancellation of this term in eq.~(\ref{+})
is the second
term on the right hand side
of eq.~(\ref{***}). The cancellation would occur iff $B^{\mu \nu}$ contained
the term of the following structure
\[
\beta^{\mu\nu}[U]\frac{\partial f}{\partial \alpha}
\]
with
\begin{equation}
\partial_\mu\beta^{\mu \nu}[U]=-\frac{1}{24 \pi^2} \epsilon^{\nu
           \sigma\lambda\rho} \mbox{Tr} \left( U\partial_\sigma U^{\dagger}
           U\partial_\lambda U^{\dagger} U\partial_\rho U^{\dagger}
           \right)\;.
\label{IV*}
\end{equation}
However, the right hand side of eq.~(\ref{IV*}) is not a complete
divergence of any tensor 
that is invariant under the flavor group (recall that
the right hand side of eq.~(\ref{IV*}) is a topological
current). Hence, the conservation of the baryonic current requires
that $\partial f/\partial\alpha=0$.

This argument is straightforward to generalize to other conserved
currents and to theories other than QCD. 
As discussed in section \ref{1}, it implies that induced 
charges should match in fundamental and low energy theories.
Examples of such a matching are given in section \ref{3}.

\section{Failure of derivative expansion: an example}
\label{2a}

An important ingredient in
the above 
argument is the derivative expansion. While in QCD and many other
models the derivative expansion works, at least to all orders of
perturbation theory, one can design models where the derivative 
expansion fails, and the induced charges cannot be reliably calculated
even within perturbation theory. As an example, let us consider a
model of free left-handed fermions $\psi^a$, $\tilde{\psi}_{\tilde{a}}$ and
$\chi^{\tilde{a}}$, $a, \tilde{a}=1,\ldots,N_f$, with the mass term
\begin{equation}
{\cal L}_{\mu_0}=\mu_0\tilde{\psi}_{\tilde{a}}\chi^{\tilde{a}}+{\rm h.c.}
\label{a4*}
\end{equation}
The model is invariant under the global $SU(N_f)_L\times SU(N_f)_R$
symmetry under which $\psi$, $\tilde{\psi}$, and $\chi$ transform as
$(N_f,1)$, $(1,\bar{N}_f)$, and $(1,N_f)$, respectively.  The ``baryon
numbers'' of fermions $\psi,\tilde{\psi}$ and $\chi$ are $+1$, $-1$ and
$+1$, respectively. Let us introduce background fields
$m^{\tilde{a}}_b(x)$ and their interaction with fermions 
$\psi$ and $\tilde{\psi}$ in the same
way as in eq.~(\ref{V*}).

\begin{figure}[htb]
\begin{center}
\begin{picture}(0,0)%
\includegraphics{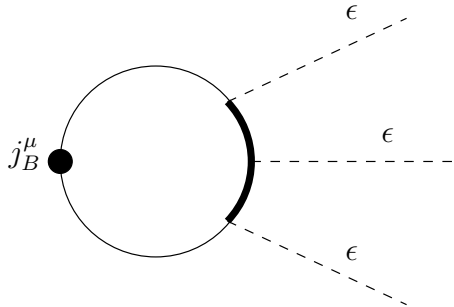}%
\end{picture}%
\setlength{\unitlength}{3947sp}%
\begingroup\makeatletter\ifx\SetFigFont\undefined%
\gdef\SetFigFont#1#2#3#4#5{%
  \reset@font\fontsize{#1}{#2pt}%
  \fontfamily{#3}\fontseries{#4}\fontshape{#5}%
  \selectfont}%
\fi\endgroup%
\begin{picture}(2712,1896)(976,-1873)
\put(3001,-1576){\makebox(0,0)[lb]
{\smash{\SetFigFont{12}{14.4}{rm}$\epsilon$}}}
\put(3001,-61){\makebox(0,0)[lb]
{\smash{\SetFigFont{12}{14.4}
{rm}$\epsilon$}}}
\put(3226,-836){\makebox(0,0)[lb]
{\smash{\SetFigFont{12}{14.4}
{rm}$\epsilon$}}}
\put(886,-961){\makebox(0,0)[lb]
{\smash{\SetFigFont{12}{14.4}{rm}$j_B^\mu$}}}
\end{picture}
\caption{Dangerous diagram  in the model
of section~\ref{2a}. Heavy and light lines correspond to massless and
massive fermions, respectively.}
\label{fig2}
\end{center}
\end{figure}

To see that the derivative expansion is not reliable in this 
model, let us again consider the background field of the form
(\ref{VI*}).  At $\epsilon=0$, fermions $\tilde{\psi}$ and \mbox{$\xi={\rm
const}\cdot(\mu_0\chi+ m_0\psi)$} form massive Dirac field, while
$\eta={\rm const}\cdot (m_0\chi-\mu_0\psi)$ 
remains  massless Weyl field. Both
types of fermions interact with the background field $\epsilon(x)$.
In an attempt to calculate the induced baryonic current, one faces
diagrams with massless internal fermion lines like one shown in 
Fig.~\ref{fig2}.
It is straightforward to see that the derivative expansion of these
diagrams is  singular.

The fact that the derivative expansion fails in this model manifests
itself in different values of induced charges in various
limits. Namely, at $\mu_0\gg m_0$ one can ignore the background
fields, and $\langle B\rangle=0$. On the other hand, at $\mu_0\ll
m_0$, the mass term (\ref{a4*}) becomes irrelevant, so $\langle
B\rangle=N[U]$.  As outlined above, this phenomenon is due to the fact
that not all fermions charged under $SU(N_f)_L\times SU(N_f)_R$ obtain
masses upon introducing the background fields $m(x)$.

This example shows that the validity of the derivative expansion requires
that the background scalar fields provide masses to all relevant
fermions. This will be the case in all examples presented in the next
section. 

\section{Examples of induced charge matching}
\label{3}

\subsection{QCD}
\label{3.1}

We again consider conventional $SU(N_c)$ QCD with $N_f$ massless
flavors. Let us generalize slightly the discussion of section~\ref{2}
by introducing background fields 
\begin{equation} 
m^{\tilde p}_{q}({\bf
x}) = m_0 U^{\tilde p}_{q}({\bf x})
\label{22*}
\end{equation} 
which give $x$-dependent masses to $N_0$ quark flavors only,
\mbox{${\cal L}_{\rm int} = m^{\tilde p}_{q}({\bf
x})\tilde{\psi}_{\tilde{p}} \psi^q \; + \; \mbox{h.c.}$}. The fields
$m({\bf x})$ are $N_0 \times N_0$ matrices; hereafter the indices
$p,q,r; \tilde{p}, \tilde{q}, \tilde{r}$ run from $1$ to $N_0$ with
$N_0 < N_f$.  $(N_f - N_0)$ flavors remain massless.  In all examples
of this section we consider background fields that are constant at
spatial infinity; by a global $SU(N_f)_L$ rotation we set $ U({\bf x})
\to {\bf 1} \;\; {\rm as} \;\; |{\bf x}| \to \infty\;.$ Besides the
baryon number symmetry $U(1)_B$, we will be interested in a vector
subgroup $U(1)_8^f$ of the original $SU(N_f)_L\times SU(N_f)_R$ flavor
group, whose (unnormalized) generator is
\[
    T_8^f = \mbox{diag}\left( 1, \dots, 1, -\frac{N_0}{N_f - N_0},\dots,
                  -\frac{N_0}{N_f - N_0}\right)\;.
\]
The background fields are charged under neither $U(1)_B$ nor
 $U(1)_8^f$. 

As all fundamental fermions that interact with the background scalar 
fields acquire masses due to this interaction, the derivative expansion
is justified, at least order by order in perturbation theory. Hence,
for slowly varying $m({\bf x})$ one has
\begin{equation}
    \langle B \rangle = N_c N[U]\;,
\label{5*}
\end{equation}  
\begin{equation}
 \langle T_8^f\rangle = N_c N[U]\;.
\label{5**}
\end{equation}
Let us see that the low energy effective theory of QCD --- the
non-linear sigma-model --- indeed reproduces eqs. (\ref{5*}) and
(\ref{5**}).  In the absence of the background fields, the non-linear
sigma-model action contains only derivative terms for the
$SU(N_f)$ matrix valued dynamical sigma-model field $V(x)$,
including the usual kinetic term and the Wess--Zumino term. The
background field $m({\bf x})$ introduces a potential term into the low
energy effective Lagrangian,
\[
           \Delta {\cal L}_{\rm eff} =
         \mbox{Tr}\left( m^{\dagger}V + V^{\dagger}m \right)\;.
\]
For slowly varying $m$, 
the effective potential is  minimized at
\begin{equation}
 V({\bf x}) =
\left(
\begin{array}{cc}
   U({\bf x})  & 0 \\
   0 &  {\bf 1}
\end{array}
\right)\;.   
\label{6*}
\end{equation}
Hence, the induced baryonic charge appears at the classical level
\cite{polychron}; as the baryonic charge of $V({\bf x})$ is equal to
its topological number $N[V]$ times $N_c$, the induced baryonic charge
is indeed given by eq.(\ref{5*}). Likewise, it follows from the
structure of the Wess--Zumino term that the $T^f_8$ current of the
configuration of the form (\ref{6*}) is (cf. ref. \cite{farhi})
\[
    j_{8,\mu}^f = \frac{N_c}{24 \pi^2}  \epsilon_{\mu\nu\lambda\rho}
           \mbox{Tr} \left( U\partial_{\nu} U^{\dagger}
                 U\partial_{\lambda} U^{\dagger}  U\partial_{\rho}
            U^{\dagger}        
            \right)\;,
\]
so the  $T^f_8$ charge of the configuration (\ref{6*})
is given by eq.(\ref{5**}).

We see that the induced charges in QCD and its low energy effective
theory match rather trivially. The way the induced charges match
becomes more interesting when low energy theories (in the absence of
background scalar fields) contain massless
fermions.

\subsection{Supersymmetric QCD}

 Let us now consider supersymmetric QCD with $N_c$ colors and $N_f$
flavors. To be specific, we discuss the case $3N_c > N_f > N_c +3$.
This theory exhibits the Seiberg duality \cite{Seiberg-du}: the
fundamental theory contains the superfields of quarks $Q^i$ and anti-quarks
$\tilde{Q}_{\tilde{j}}$, while its effective low energy
counterpart at the origin of moduli space is an $SU(N_f - N_c)$ 
magnetic gauge theory with magnetic quarks $q_i$, magnetic anti-quarks
$\tilde{q}^{\tilde{j}}$ and mesons $M^i_{\tilde{j}}$ with the 
superpotential $qM \tilde{q}$. 

Let us probe this theory by adding the scalar background fields
$m^{\tilde{q}}_p({\bf x})$ with the same properties as above, i.e.,
by introducing the term
\begin{equation} 
   m^{\tilde{q}}_p({\bf x}) \tilde{Q}_{\tilde{q}} Q^p    
\label{31*} 
\end{equation}
into the superpotential of the fundamental theory. Let us take for 
definiteness\\ 
$2 \leq N_0 < N_f - N_c - 1$. The calculation of the induced 
baryon and $T_8^f$ charges in the fundamental theory proceeds as above,
and we again obtain eqs.(\ref{5*}) and (\ref{5**}).

We now turn  to the effective low energy theory. For slowly varying
$m({\bf x})$, the term (\ref{31*}) translates into $\mbox{Tr}(mM)$,
so the total superpotential of the magnetic theory is
\begin{equation} 
   qM\tilde{q} + \mu_0\mbox{Tr}(mM)\;,
\label{32+} 
\end{equation} 
where $\mu_0$ is the dimensionfull parameter inherent in the magnetic 
theory.
The ground state near the origin of the moduli 
space has the following non-vanishing ${\bf x}$-dependent
expectation values\footnote{Hereafter we use the same notations
for superfields and their scalar components.} of the magnetic 
quarks and anti-quarks,
\begin{equation} 
     \langle q^p_q\rangle \;= \mu_q^p\;, \;\;p= 1, \dots, N_0, \;\;
     q=1, \dots, N_0
\label{32*}
\end{equation} 
(here the upper and lower indices refer to magnetic color and
flavor, respectively) 
\begin{equation} 
    \langle \tilde{q}^{\tilde{q}}_p\rangle \; =
   \tilde{\mu}^{\tilde q}_p\;, \;\;p= 1, \dots, N_0, \;\; 
     \tilde{q}=1, \dots, N_0
\label{32**} 
\end{equation}
(here the lower index refers to magnetic color). 
The expectation values obey
\[
    \tilde{\mu}^{\tilde r}_p ({\bf x}) \mu_q^p ({\bf x}) = 
        - \mu_0 m_q^{\tilde{r}} ({\bf x})\;.
\]
They also satisfy the $D$-flatness condition at each point in space, 
$         \mu^{\dagger q}_p\mu_q^r  =
 \tilde{\mu}^{\tilde q}_p \tilde{\mu}^{\dagger r}_{\tilde q}$.
With our choice of background fields, eq.(\ref{22*}), one has
\[
     \mu = \pm \sqrt{\mu_0 m_0}~W({\bf x})\; , \;\;
     \tilde{\mu} = \mp \sqrt{\mu_0 m_0}~\tilde{W}({\bf x})\;,
\]
where $W$ and $\tilde{W}$ are $N_0 \times N_0$ unitary 
matrices
obeying
\begin{equation} 
      (\tilde{W}W)({\bf x}) = U({\bf x})\;.
\label{23++} 
\end{equation} 
Indeed, at $m = m_0\cdot {\bf 1}$, the matrices $\mu$ and 
$\tilde{\mu}$ are proportional to  $N_0 \times N_0$ unit matrix, up 
to magnetic color rotation. At $m = m_0 U({\bf x})$ one has
$\mu = \pm \sqrt{\mu_0 m_0}~U_c U$, 
$\tilde{\mu} = \mp \sqrt{\mu_0 m_0}~U_c^{\dagger}$
where $U_c({\bf x})$ is a slowly varying matrix belonging to
$SU(N_0)$ subgroup of the magnetic color group. The explicit form of
$U_c({\bf x})$ is to be found from the minimization of the gradient 
energy, and it is not important for our purposes.

Since the gradient energy has to vanish at spatial 
infinity, $W({\bf x})$ and  $\tilde{W}({\bf x})$ are constant at
$|{\bf x}| \to \infty$, so they can be characterized by their
winding numbers $N[W]$ and  $N[\tilde{W}]$. Because of eq. (\ref{23++})
one has
\[
      N[W] + N[\tilde{W}] = N[U]\;.
\]
In this ground state, the magnetic color is broken down to
$SU(N_f - N_c - N_0)$. At small $m_0$, the ground state
(\ref{32*}), (\ref{32**}) is close to the origin of the moduli
space, so the magnetic description is  reliable.

Both the baryon number and $T_8^f$ are broken in this vacuum.
However, there exist combinations of these generators and
magnetic color generators that remain unbroken.
Recalling \cite{Seiberg-du} that the baryon number of magnetic quarks
equals $N_c/(N_f - N_c)$ and that the magnetic quarks and anti-quarks
transform as $(\bar{N}_f, 1)$ and $(1,N_f)$, respectively,
under the global $SU(N_f)_L\times SU(N_f)_R$ group, 
the unbroken generators are
\begin{equation} 
     B' = B - \frac{N_c}{N_f - N_c} T_8^{mc}\;,
\label{33*}
\end{equation}
 \begin{equation} 
     T'_8 = T_8^f + T_8^{mc}\;,
\label{33**}
\end{equation}
where $ T_8^{mc}$ is the following generator of the magnetic color group,
\[
    T_8^{mc} = \mbox{diag} \left(1,\dots,1, -\frac{N_0}{N_f - N_c - N_0},
    \dots -\frac{N_0}{N_f - N_c - N_0} \right)\;.
\]
As the fundamental quarks and gluons are singlets under magnetic
color, the induced charges $\langle B'\rangle$ and $\langle
T'_8\rangle$ calculated in the magnetic theory should match
eqs. (\ref{5*}) and (\ref{5**}). Let us check that this is indeed the
case.

The induced charges appear in the magnetic theory through ${\bf
x}$-dependent mass terms of fermions. These are generated by the
expectation values (\ref{32*}), (\ref{32**}). The mass terms coming
from the superpotential (\ref{32+}) are
\begin{equation} 
    \tilde{\mu}^{\tilde{p}}_q ({\bf x})
\Psi_{\tilde{p}}^i \psi_i^q
   + \mu^{p}_q ({\bf x})
\tilde{\psi}^{\tilde{j}}_p \Psi_{\tilde{j}}^q\;,
\label{44**} 
\end{equation}
where $\Psi$, $\psi$ and $\tilde{\psi}$ are fermionic components
of mesons, magnetic quarks and magnetic anti-quarks, respectively.
The gauge interactions give rise to other mass terms,
\begin{equation} 
   \mu^{\dagger p}_q ({\bf x})
\psi^a_p \lambda^q_a
   - \tilde{\mu}^{\dagger p}_{\tilde{q}}  ({\bf x})
\lambda^a_p 
      \tilde{\psi}_a^{\tilde q}\;, 
\label{44*} 
\end{equation}
where $\lambda^b_a$ is the gluino field, $a,b = 1,\dots, (N_f - N_c)$
are magnetic color indices.

To calculate the induced baryon number $\langle B'\rangle$ we observe
that the only fermions carrying non-zero $B'$ are magnetic quarks
$\psi_i^{\alpha}$ with $i = 1,\dots, N_f$, \\ $\alpha = (N_f - N_c -
N_0 + 1), \dots, (N_f - N_c)$, magnetic anti-quarks
$\tilde{\psi}^{\tilde{j}}_{\alpha}$ and gluinos $\lambda^{\alpha}_p$,
$\lambda_{\alpha}^p$. Their $B'$-charges are
\[
   \psi_i^{\alpha}: \; \frac{N_c}{N_f - N_c} -
      \frac{N_c}{N_f - N_c}\left(-\frac{N_0}{N_f - N_c - N_0}  
      \right) = \frac{N_c}{N_f - N_c - N_0}\;,
\]
\[
   \lambda^{\alpha}_p : \; \frac{N_c}{N_f - N_c - N_0}\;,
\]
\[
  \tilde{\psi}^{\tilde{j}}_{\alpha}\;, \;\; \lambda_{\alpha}^p
     : \; -\frac{N_c}{N_f - N_c - N_0}\;.
\]
Hence, the induced $B'$ is due to the ${\bf x}$-dependent mass term
(\ref{44*}) and is equal to
\[
  \langle B'\rangle =  -\frac{N_c}{N_f - N_c - N_0} \cdot (N_f - N_c - N_0)
         \left( N[W^{\dagger}] +  N[\tilde{W}^{\dagger}] \right)\;.
\]
It follows from eq. (\ref{23++}) that $\langle B'\rangle$ indeed
coincides with $N_c N[U]$, the induced baryon number calculated in the
fundamental theory.

The induced charge $\langle T'_8\rangle$ is calculated in a similar way.
The relevant $T'_8$ charges of magnetic quarks are
\[
   \psi_p^{\alpha}: \; -\frac{N_f - N_c}{N_f - N_c - N_0}\;,
\]
\[
\psi_p^{u}: \; \frac{N_f}{N_f - N_0}\;, \;\; u = (N_0 +1), \dots, N_f\;,
\]
and similarly for magnetic anti-quarks, gluinos and mesons.
We find that both ${\bf x}$-dependent mass terms, (\ref{44**}) and
(\ref{44*}), contribute to $\langle T'_8\rangle$, and obtain
\begin{eqnarray}
\langle T'_8\rangle& =&  \frac{N_f}{N_f - N_0}\cdot (N_f - N_0)
         \left( N[W] + N[\tilde{W}] \right)\nonumber\\
    &+&
       \frac{N_f - N_c}{N_f - N_c - N_0} \cdot (N_f - N_c - N_0)
        \left( N[W^{\dagger}] + N[\tilde{W}^{\dagger}] \right)\;.
\nonumber
\end{eqnarray}
This is equal to $N_c N[U]$, so the induced $T'_8$ charges also
match in the fundamental and low energy theories.

\subsection{Softly broken SQCD}

 As our last example, let us consider supersymmetric QCD with small
soft masses of scalar quarks, $m_Q^2$, that explicitly break
supersymmetry \cite{Peskin}. We again probe this theory by introducing
the term (\ref{31*}) into the superpotential. The restrictions on
$N_f$, $N_c$ and $N_0$ are the same as in the previous example.

The induced charges, as calculated in the fundamental theory,
are still given by eqs. (\ref{5*}) and (\ref{5**}). The low energy 
theory near the origin is still the magnetic theory, but now
with soft mass terms of scalar mesons and scalar magnetic quarks
\cite{Peskin}. The scalar potential of the magnetic theory near the 
origin
at small $m_Q^2$ is determined both by the superpotential (\ref{32+}) 
and these soft terms,
\begin{equation} 
   V(M, q, \tilde{q}) = |\tilde{q}q + \mu_0 m|^2 + |qM|^2
                       + |M\tilde{q}|^2 + m_M^2 M^{\dagger} M
       + m_q^2 (q^{\dagger}q + \tilde{q}^{\dagger} \tilde{q})
       + D\mbox{-terms}\;,
\label{52*} 
\end{equation}
where $m_M^2$ and $m_q^2$ are proportional to $m_Q^2$. Were the soft
terms in eq. (\ref{52*}) positive, the ground state of this theory at
$m_q^2 > \mu_0 m_0$ would be at the origin, $\langle q\rangle =
\langle \tilde{q}\rangle = \langle M\rangle =0$. The masses of
fermions in the magnetic theory would vanish, the induced charges
$\langle B\rangle$ and $\langle T_8^f\rangle$ would be zero, so the
induced charge matching would not occur. Hence, the induced charge
matching {\it requires} that either $m_q^2$ and/or $m_M^2$ are
negative, so that the ground state even at $m_0=0$ is far 
away from the
origin, or $m_q^2 = 0$, $m_M^2 \geq 0$ with the ground state being the
same as in the previous example. This is in accord with explicit
calculations. Indeed, it has been found in ref. \cite{Rattazzi} (see also
ref. \cite{another}) that $m_q^2 < 0$ at $N_c + 1 < N_f < 3N_c/2$,
i.e., when the magnetic theory is weakly coupled. 
As regards
the conformal 
window
 $3N_c/2 \leq N_f < 3N_c$,
the complete analysis of the infrared soft masses is
still lacking. However, the existing results \cite{conformal}
(see also ref. \cite{Rattazzi}) suggest that either the origin
$q = \tilde{q} = M = 0$ is again unstable, or 
$m_q^2 = m^2_M = 0$. We
conclude that the induced charge matching provides qualitative
understanding of the behavior of soft masses in low energy description
of softly broken SQCD.

One can show that the same phenomenon
occurs in softly broken supersymmetric theories with $SO(N_c)$ and 
$Sp(2k)$ gauge groups and fundamental quarks at $N_f$, $N_c$
and $k$ such that the Seiberg duality holds. In particular,
in theories with weakly coupled magnetic description, the soft masses 
of magnetic squarks are negative. 
This is again in full accord with induced charge matching 
conditions.

It is worth noting that there exists an example \cite{Rattazzi} where
soft masses of fundamental scalar quarks single out the vacuum at the
origin of the moduli space (in the absence of the background fields $m
({\bf x})$). This is the theory with $Sp(2k)$ gauge group and $2k +4 =
2N_f$ quarks $Q_i$, $i=1,\dots, 2N_f$, in the fundamental
representation. The low energy effective theory \cite{sp2k} contains
antisymmetric mesons $M_{ij}$ and has superpotential
$\mbox{Pf}~M$. One can probe this theory by adding ${\bf x}$-dependent
mass terms $m^{p\tilde{q}}({\bf x}) Q_{\tilde{q}}Q_p$ where
$p=1,\dots, N_f$, $\tilde{q} = (N_f+1),\dots, 2N_f$. In the theory
without soft supersymmetry breaking, the induced charges match in a
similar way as in the previous example: scalar mesons obtain the
expectation values  $\langle M_{\tilde{q}p} ({\bf x})\rangle \propto
m^{\dagger}_{\tilde{q}p} ({\bf x})$ which give ${\bf x}$-dependent
masses to fermionic mesons. After the soft scalar quark masses are
introduced, the scalar potential of the low energy theory contains
soft meson masses, $m_M^2 M^{\dagger}M$ where $m_M^2 > 0$ at $k>1$
\cite{Rattazzi}. At first sight, this ruins the induced charge
matching at small $m_0$, as the ground state appears to be at \mbox{$M=0$}
and no ${\bf x}$-dependent masses of fermionic mesons seem to be
generated. However, the symmetries of the theory allow for a linear
supersymmetry breaking term in the scalar potential, $m_Q^2 f(m_Q^2,
m_0) mM$, which shifts the ground state to $\langle M\rangle \propto
m^{\dagger}$ and in this way restores induced charge matching. Hence,
we argue that this linear term is indeed generated in the low energy
theory.

\section{Discussion}
\label{4}

Let us discuss the relation between induced charges and triangle
anomalies; we consider induced baryon number in QCD as an example. We
use the notations of section \ref{3.1}.  The ${\bf x}$-dependence of
the background field $m({\bf x})$ can be removed at the expence of
modification of the gradient term in the quark Lagrangian.  Namely,
after the $SU(N_0)_L$ rotation of the left-handed quark fields, $\psi
(x) \to U^{-1}({\bf x}) \psi (x)$, $\tilde{\psi}(x) \to
\tilde{\psi}(x)$, first $N_0$ quarks and anti-quarks have ${\bf
x}$-independent masses $m_0$, and the gradient term of these quarks
becomes $\bar{\psi}~ i \gamma \cdot\left(D + \frac{1- \gamma^5}{2}
A^L\right) \psi$ where $A_0^L =0$, $A^L_i = U \partial_iU^{-1}$, the
covariant derivative $D_{\mu} $ contains dynamical gluon fields, and
we  switched to four-component notations.  The addition to
the gradient term may be viewed as the interaction of massive quarks
with the background pure gauge vector fields corresponding to
$SU(N_0)_L$ subgroup of the flavor group; these background fields are
small, time-independent, and slowly varying in space.

Now, consider an adiabatic process (either in Minkowskian
or in Euclidean space-time) in which the background vector
fields ${\bf A}^L (x)$  (in the gauge $A^L_0 = 0$) change 
in time from $ A^L_i = 0$
to $A^L_i = U \partial_iU^{-1}$, always varying slowly 
in space and vanishing at spatial infinity (an example of such a 
process is an instanton of large size).
 Suppose that this process 
begins with the system in the ground state which has zero induced 
charges because of the triviality of the background.  As the 
background vector fields interact with massive degrees of freedom only,
the system remains in its ground state in the entire process, at least
order by order in perturbation theory. The induced baryon number in the
final state --- the quantity we are interested in --- is equal to
$\langle B\rangle = \int~d^4 x~ \partial_{\mu} j^{\mu}_B$ which in turn is
determined by the anomaly in the gauge-invariant baryonic current 
$j^{\mu}_B$. Hence, we recover eq. (\ref{5*}):
\begin{equation}
\langle B\rangle = -\frac{N_c}{16 \pi^2} \int\!d^4 x~{\rm Tr} F^L_{\mu\nu}
         \tilde{F}^L_{\mu\nu} 
       = N_c N[U]\;.
\label{11*}
\end{equation}
This observation relates induced charges and anomalies.

Let us note in passing that the phenomenon discussed in section
\ref{2a} may be understood in this language as follows.  After the
background field $m(x)$ is introduced, and its phase $U({\bf x})$ is
rotated away, there remain {\it massless} fermions interacting with
the background field ${\bf A}^{L}({\bf x})$. In this situation the adiabatic
process does not necessarily end up in the ground state, because some
energy levels of massless fermions may cross zero. In that case the
baryon number induced in the ground state by the background field
$A^L_i ({\bf x}) = U \partial_i U^{-1} ({\bf x})$ would be different
from eq. (\ref{11*}), as the anomaly determines the {\it total} change
in the baryon number. This is the reason for the dependence of the
induced charges on the parameters of the theory ($m_0/\mu_0$ in the
example of section \ref{2a}).

One may wonder whether similar phenomenon (fermion level crossing)
might occur even if all relevant fundamental
fermions obtain masses upon
introducing the background scalar fields, i.e.,
whether the final state might actually contain excitations carrying
non-zero net baryon number. To argue that this does not happen, 
let us consider QCD again. 
The appearance, in the final state, of excitations with
non-zero net baryon number would show up as a non-vanishing index
of the four-dimensional
Euclidean Dirac operator ${\cal D}[A^L] = \gamma 
\cdot \left(D + \frac{1-\gamma^5}{2} A^L (x) \right) + m_0$, so
that the vacuum-to-vacuum amplitude would vanish while matrix elements
of baryon number violating operators between the 
initial and final vacua
would not. However, for arbitrary gluon fields, the 
eigenvalues $\omega$ of the operator ${\cal D}[A^L = 0] = \gamma 
\cdot D + m_0$ obey $|\omega| > m_0 $ (the Euclidean operator
$\gamma \cdot D $ is anti-Hermitean) so the operator ${\cal D}[A^L]$
has no zero modes when the background fields $A^L (x)$ are small
($A^L (x) \ll m_0$ at all $x$) and slowly vary in space-time.
This argument implies that eq. (\ref{11*}) is valid in full quantum 
theory even at $m_0 < \Lambda_{QCD}$. Although the situation in
theories with fundamental scalars is more complicated, it is likely 
that analogous arguments may be designed in those theories as well.

Finally, 
the same
adiabatic process may be considered 
within the low energy effective theory.
The induced baryon number is now related to the anomaly in the
effective theory,  {\it provided}
all low energy degrees of freedom interacting with $SU(N_0)_L$
gauge fields become massive upon introducing the mass $m_0$
to $N_0$ flavors of fundamental quarks. As the 
$U(1)_B \times SU(N_0)_L \times SU(N_0)_R$ anomalies are the
same in the fundamental and low energy theories,
the induced baryon numbers match automatically in that 
case. 
This observation has an obvious generalization: 
a sufficient condition for induced charge matching is that
no low energy degrees of freedom transforming non-trivially under
a subgroup of the flavor group remain massless when this subgroup is
explicitly broken by masses of some fermions of the fundamental 
theory. This property is certainly valid in supersymmetric theories
where no phase transition is expected to occur as the masses of some of
the flavors flow from small to large values, i.e., where
massive flavors smoothly decouple. On the other hand, this property
does not seem to be guaranteed in non-supersymmetric models,
though it is intuitively appealing and may well be quite generic.

 The authors are indebted to E.~Akhmedov, D.~Amati, G. Dvali,
Yu.~Kubyshin, A.~Ku\-zne\-tsov, V.~Kuzmin, A.~Penin, K.~Selivanov,
A.~Smirnov, P.~Tinyakov and S.~Troitsky for numerous helpful
discussions. This research was supported in part by Russian Foundation
for Basic Research under grant 99-02-18410.  The work of S.D. and
D.G. is supported in part by INTAS under grant 96-0457 within the
research program of the International Center for Fundamental Physics
in Moscow and by ISSEP fellowships.  The work of S.D., D.G., and M.L
was supported in part by the Russian Academy of Sciences, JRP grant
37.  V.R. would like to thank Professor Miguel Virasoro for
hopsitality at the Abdus Salam International Center for Theoretical
Physics, where part of this work was carried out.

\def\ijmp#1#2#3{{\it Int. Jour. Mod. Phys. }{\bf #1~}(19#2)~#3}
\def\pl#1#2#3{{\it Phys. Lett. }{\bf B#1~}(19#2)~#3}
\def\zp#1#2#3{{\it Z. Phys. }{\bf C#1~}(19#2)~#3} 
\def\prl#1#2#3{{\it Phys. Rev. Lett. }{\bf #1~}(19#2)~#3} 
\def\rmp#1#2#3{{\it Rev. Mod. Phys. }{\bf #1~}(19#2)~#3} 
\def\prep#1#2#3{{\it Phys. Rep.}{\bf #1~}(19#2)~#3} 
\def\pr#1#2#3{{\it Phys. Rev. }{\bf D#1~}(19#2)~#3} 
\def\np#1#2#3{{\it Nucl. Phys. }{\bf B#1~}(19#2)~#3} 
\def\mpl#1#2#3{{\it Mod. Phys. Lett. }{\bf #1~}(19#2)~#3} 
\def\arnps#1#2#3{{\it Annu. Rev. Nucl. Part. Sci.}{\bf #1~}(19#2)~#3} 
\def\sjnp#1#2#3{{\it Sov. J. Nucl. Phys.}{\bf #1~}(19#2)~#3} 
\def\jetp#1#2#3{{\it JETP Lett. }{\bf #1~}(19#2)~#3} 
\def\app#1#2#3{{\it Acta Phys. Polon. }{\bf #1~}(19#2)~#3} 
\def\rnc#1#2#3{{\it Riv. Nuovo Cim. }{\bf #1~}(19#2)~#3} 
\def\ap#1#2#3{{\it Ann. Phys. }{\bf #1~}(19#2)~#3}
\def\ptp#1#2#3{{\it Prog. Theor. Phys. }{\bf #1~}(19#2)~#3}
\def\spu#1#2#3{{\it Sov. Phys. Usp.}{\bf #1~}(19#2)~#3}

\end{document}